\pdfoutput=1

\documentclass[11pt]{article}

\usepackage[final]{acl}
\usepackage{graphicx}
\usepackage{multirow}
\usepackage{times}
\usepackage{latexsym}

\usepackage[T1]{fontenc}

\usepackage[utf8]{inputenc}

\usepackage{microtype}

\usepackage{inconsolata}

\usepackage[normalem]{ulem}
\usepackage{algorithm}
\usepackage{multirow}
\usepackage{amsmath} 
\usepackage{enumitem}

\usepackage{amsmath}
\usepackage{booktabs}

\usepackage{algorithm}
\usepackage{algpseudocode}
\usepackage[most]{tcolorbox}

\title{STARD: A Chinese Statute Retrieval Dataset with Real Queries Issued by Non-professionals}

\usepackage{authblk}
\author[1]{\textbf{Weihang Su}\thanks{swh22@mails.tsinghua.edu.cn}}
\author[2]{\textbf{Yiran Hu}\thanks{Equal Contribution.}}
\author[3]{\textbf{Anzhe Xie}}
\author[1]{\textbf{Qingyao Ai}\thanks{Corresponding Author: aiqy@tsinghua.edu.cn}}
\author[2]{\textbf{Zibing Que}}
\author[4]{\textbf{Ning Zheng}}
\author[2]{\textbf{Yun Liu}} 
\author[2]{\\ \textbf{Weixing Shen}}
\author[1]{\textbf{Yiqun Liu}}

\affil[1]{Department of Computer Science and Technology, Tsinghua University}
\affil[2]{School of Law, Tsinghua University}
\affil[3]{School of Electronics Engineering and Computer Science, Peking University}
\affil[4]{Department of Automation, Tsinghua University}

\begin{document}
\maketitle

\begin{abstract}

Statute retrieval aims to find relevant statutory articles for specific queries. This process is the basis of a wide range of legal applications such as legal advice, automated judicial decisions, legal document drafting, etc.
Existing statute retrieval benchmarks focus on formal and professional queries from sources like bar exams and legal case documents, thereby neglecting non-professional queries from the general public, which often lack precise legal terminology and references. 
To address this gap, we introduce the STAtute Retrieval Dataset (STARD), a Chinese dataset comprising 1,543 query cases collected from real-world legal consultations and 55,348 candidate statutory articles\footnote{All the codes and datasets are available at: https://github.com/oneal2000/STARD/tree/main}.
Unlike existing statute retrieval datasets, which primarily focus on professional legal queries, STARD captures the complexity and diversity of real queries from the general public.
Through a comprehensive evaluation of various retrieval baselines, we reveal that existing retrieval approaches all fall short of these real queries issued by non-professional users. The best method only achieves a Recall@100 of 0.907, suggesting the necessity for further exploration and additional research in this area.



\end{abstract}

\section{Introduction}

\begin{table}[t]
\centering
{\small{\begin{tabular}{|p{0.92\columnwidth}|}
\toprule

\textbf{Question:} Is disclosing the medical case information of a patient considered an invasion of privacy?\\

\midrule

\textbf{Relevant Statute Articles}\\
\textbf{Personal Information Protection Law, Article 28:} Sensitive personal information refers to information that, if leaked or illegally used, could easily harm an individual's dignity or endanger their personal or property safety. This includes biometric data, religious beliefs, specific identities, medical health, financial accounts, tracking information, and personal information of minors under the age of fourteen.\\

\midrule

\textbf{Civil Code, Article 1032:} Individuals have the right to privacy. No organization or individual may infringe upon another's privacy rights through snooping, harassment, disclosure, or publicization. Privacy encompasses the tranquility of an individual's private life and the private spaces, activities, and information they wish to keep unknown to others.\\

\midrule

\textbf{Civil Code, Article 1226:} Medical institutions and their medical personnel must keep patients' privacy and personal information confidential. Those who disclose patients' private and personal information or publish their medical records without the patient's consent must bear infringement liability.\\

\toprule
\end{tabular}}}
\vspace{-3mm}
\caption{An example of the query and relevant statute articles in the STARD dataset.}
\label{tab:example}
\vspace{-6mm}
\end{table}

Statutes are written laws formally created and approved by a legislative body, such as a parliament or congress~\cite{livingston1990congress}. 
They set out specific rules and guidelines within a certain area or jurisdiction. 
Therefore, statutes are the primary source of legal authority in civil law countries and also play a significant role in common law jurisdictions.



Statute retrieval involves finding relevant statutory articles or sections of laws for a specific query. 
This process is vital in the legal field and supports a wide range of applications, including legal advice services, automated judicial decisions, and logical legal analysis. 
This task is challenging for the following reasons:
\noindent \textbf{(1)} 
Statutes use complex legal terminology and linguistic structures rarely found in open-domain corpus. As a result, traditional retrieval models that lack domain-specific knowledge may struggle to accurately capture the meanings of these specialized terms.
\noindent \textbf{(2)} 
The criteria for assessing information relevance in the legal domain differ greatly from those used in open-domain search tasks. 
General search tasks focus mainly on textual similarity, while legal tasks involve legal reasoning that requires the understanding of different areas of law, the relations between them, as well as the relevance of specific legal principles and their practical applications.


Due to the challenging nature of statute retrieval and its paramount importance in civil law systems, significant progress has been made in this field. 
For example, the annual COLIEE competitions introduce a series of statute retrieval tasks using the questions extracted from the Japanese legal bar exams~\cite{goebel2023summary,kim2022coliee,rabelo2022overview}. 
These tasks aim to retrieve relevant statute law from the Japanese Civil Code Article according to the question from bar exams.
AILA~\cite{bhattacharya2019fire} competitions also introduce a series of statute retrieval datasets. 
The queries from AILA are case 
documents that were judged by the Supreme Court of India. 
The candidate statutes are part of the set of statutes from Indian law.

Despite these advancements, a significant gap persists in addressing real queries from non-professional people, who represent a large population of legal advice service users. 
The current statute retrieval benchmarks are primarily based on queries from formal legal documents, such as bar exam questions or Supreme Court case documents, which differ significantly from the everyday language used by the general public. 
However, queries from non-professionals often lack precise legal terminology and may include ambiguous references to legal concepts, which significantly complicate the task of statute retrieval.

To address the limitations of existing benchmarks, 
we propose STAtute Retrieval Dataset (STARD) i.e. STARD, a Chinese statute retrieval dataset based on real legal consultation questions from the general public. 
The STARD dataset comprises 1,543 query cases collected from real-world legal consultations and 55,348 candidate statutory articles extracted from all official Chinese legal regulations and judicial interpretations.  
Table~\ref{tab:example} shows an example of our dataset.
To the best of our knowledge, STARD is the first statute retrieval dataset where queries are from real-world legal consulting proposed by the general public.

We conduct experiments on a wide range of information retrieval (IR) baselines on the STARD dataset, including traditional lexical matching models, open-domain neural retrieval models, legal domain neural retrieval models, and a dense retriever trained with data annotated by GPT-4. The experimental results show that all existing baselines fall short of accurately and comprehensively retrieving the relevant statutes, leaving significant room for future work.
Additionally, our experimental results show that employing STARD as an external knowledge source for Retrieval-Augmented Generation (RAG) significantly enhances the performance of large generative language models (LLMs) on legal tasks. This indicates that STARD is useful for developing more accessible and efficient legal systems.

In conclusion, the contributions of this paper are as follows:

\begin{itemize}[leftmargin=*]
    \item We propose STARD, a statute retrieval dataset derived from real-world legal consultation posed by non-professionals, with 1,543 queries and their corresponding relevant statutes.  

    \item We propose a comprehensive annotation framework specifically designed for the statute retrieval task based on non-professional queries, which provides references and insights for future annotation in the legal field.

    \item We conduct experiments on a wide range of retrieval baselines and find that statute retrieval with queries issued by non-professionals is still a difficult task that requires further investigation. 

    \item We present experiments on LLMs solving legal tasks with and without the STARD dataset. 
    Experiments show that STARD can notably enhance the performance of LLMs in legal tasks.
    
\end{itemize}

\section{Problem Formulation}
\subsection{Statute of Civil Law System}

Civil law is a legal system primarily based on codified laws rather than case precedents, making written statutes the main source of legal authority. This contrasts with common law systems, where previous judicial decisions also play a central role. 
In civil law, statutes are created and enacted by legislative bodies, such as parliaments, and are organized into systematic collections known as codes, which cover various areas of law like contracts, torts, and property. 
A statute is a formal written law that provides specific rules and guidelines to be followed within a jurisdiction. Within statutes, there are sections known as statutory articles, which detail individual provisions or clauses of the law, addressing particular aspects or requirements. These statutes and their articles are fundamental in civil law systems to ensure that the legal framework is clear, predictable, and accessible, thereby facilitating order and defining societal rights and responsibilities.

\begin{figure*}[ht]
  \centering
  \includegraphics[width=\textwidth]{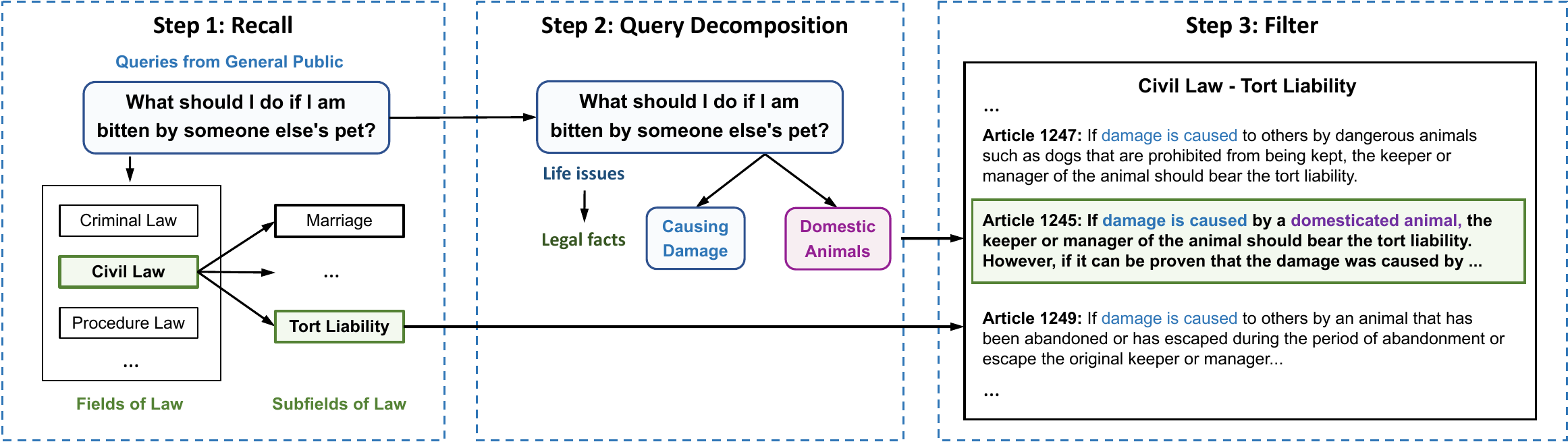}
  \caption{A schematic of our annotation framework with green boxes highlighting query-relevant elements.} 
  \label{fig:framework} 
\end{figure*}

\subsection{Definition of Statute Retrieval}
The statute retrieval task aims to accurately retrieve relevant statutory articles in response to a query. 
To be specific, given a query $q$ that describes a legal issue or situation and a corpus of statutory articles $S=\{s_1, s_2, ..., s_n\},$ $n \in N^+$. 
For each statute $s_i$ in the corpus, there is a Bernoulli variable $r_i$ indicating whether $s_i$ is relevant\footnote{The definition of ``relevant'' is discussed in detail in Section~\ref{sec:relevance}.} to the query $q$. 
The goal of the statute retrieval task is to retrieve a set of statutes $R=\{s_j | r_j=1\}$, including all statutes relevant to the query. 



\section{Annotation Framework}
\label{sec:relevance}

This section explains how annotators transform general questions into professional legal questions submitted by non-professionals and then identify the most relevant legal statutes to support these questions. 
To be specific, annotators use a three-step method: recall, query decomposition, and filtering (illustrated in Figure~\ref{fig:framework}). This method mirrors the structured approach commonly used in legal reasoning, which involves three logical steps: establishing a broad legal principle (major premise), applying it to the specific facts of a case (minor premise), and then concluding. 
This section is organized into three subsections, each detailing a part of the annotation process that is designed to mirror these logical steps in legal reasoning.


\subsection{Step 1: Recall}

When initiating the annotation of legal statutes pertinent to a query, our annotators first narrow down the scope of the relevant statutes.
Specifically, they start by identifying the most pertinent areas of law within the entire legal system.
The process uses a top-down refining method. Annotators begin with broad departmental categories of law, such as civil, criminal law, and administrative law.
Upon encountering a specific issue, annotators first determine which category of departmental law it falls under, then progressively refine the issue to more specific aspects of the law. 
For instance, if the issue pertains to civil law, the annotator assesses whether it relates to contract law or tort law. If it is a matter of contract law, a further determination is made regarding the specific type of contract involved. Similarly, for tort law, the specific type of tort is identified. This step effectively narrows the scope of legal statute retrieval to particular chapters within the relevant departmental law.

\subsection{Step 2: Query Decomposition}
Given the specialized nature of legal knowledge, individuals without a formal education in law often frame their queries with informal language rather than professional legal terminology. 
These queries typically consist of straightforward semantic expressions that do not directly correspond to established legal norms.
For instance, consider the question ``What should I do if I am bitten by someone's pet?''. Here, ``pet bite'' represents a common, non-technical description of an incident. Searching for legal norms based solely on such descriptions might lead to irrelevant or imprecise results.
Therefore, when annotators perform legal statute retrieval, they should transform the informal fact descriptions written by the questioner into legal facts through interpretation. This is the step to find the minor premise in the legal logic syllogism. In this transformation process, the annotator evaluates the life facts according to the provisions of the law and selects the legal norms corresponding to these life facts. For example, for the aforementioned issue of a pet biting a person, the annotators will transform ``pet bites a person'' into the legal fact of ``causing damage to other'' and ``domestic animals'' according to the provisions of Chapter 9 of the Tort Liability Compilation of the Civil Code.

\subsection{Step 3: Filter}

The filtering process is a critical step where annotators refine and finalize the selection of relevant legal statutes. This is accomplished by employing a ``subsumption'' method, integral to the syllogistic reasoning in law. In this method, the legal facts, which have been interpreted and transformed from real-life scenarios in the previous steps, are matched against the smallest possible subset of legal statutes that adequately address the query.

To be specific, consider a set of legal statutes \( S = \{S_1, S_2, S_3\} \) recalled in the first step. Through the transformation process, the query is deconstructed into distinct legal facts \( F_1, F_2, F_3 \). Each fact corresponds to a subset of statutes that it implies, denoted as \( S_{F_1} = \{S_1, S_4, S_5\} \), \( S_{F_2} = \{S_1, S_5\} \), and \( S_{F_3} = \{S_3, S_6\} \). The objective in the filtering stage is to intersect these subsets with the initially recalled set \( S \) to determine the most relevant statutes. This is represented as \( S_{Golden} = (S_{F_1} \cup S_{F_2} \cup S_{F_3}) \cap S \), yielding \( S_{Golden} = \{S_1, S_3\} \).

These statutes in \( S_{Golden} \) are considered the ``golden'' legal statutes for the dataset, as they encompass all the legal implications drawn from the facts of the query. This step ensures that the selected statutes are not only relevant but also comprehensive in covering the legal issues presented in the query, thereby providing a solid legal foundation to support the resolution of the query.

\subsection{Generalizability of Our Framework}

In this section, we discuss the generalizability of the STARD dataset and our annotation framework, discussing the following two research questions (RQs):
\begin{itemize}[leftmargin=*]
\item   \textbf{RQ1:} Can the STARD dataset be applied to the legal systems of other countries through direct translation of our dataset?
\item   \textbf{RQ2:} Can our Three-Step Annotation Framework be applied to other legal systems?
\end{itemize}

For RQ1, directly translating the STARD dataset into other languages does not guarantee its applicability in foreign legal systems. Each country possesses unique legal statutes; articles selected from one jurisdiction may not exist or may have entirely different implications in another. Thus, the nuances of local laws must be considered, making straightforward translation inadequate for cross-national applications.
For RQ2, our proposed Three-Step Annotation Framework is potentially generalizable to other countries under the civil law system. Countries with civil law systems, such as Germany, France, and Japan, typically share a similar process for retrieving law statutes. This process can generally be structured into three steps: Recall, Query Decomposition, and Filtering. 
Therefore, our framework could be adapted to these environments, supporting the construction of statute retrieval datasets and the application of legal statutes across various civil law jurisdictions.

\section{Dataset Construction}

\subsection{Data Sources}

All queries in our dataset derive from real legal consultations. 
Specifically, our legal team creates legal questions from the 12348 China Legal Service Website\footnote{This is the Chinese government's official website for online legal services: http://www.12348.gov.cn/}, followed by a manual anonymization of each question, which involved removing any potential identifiers associated with entities, corporations, or individuals.

To obtain the 55,348 candidate statutory articles, our legal team conducted extensive research and discussions to compile a comprehensive list of currently valid Chinese statutory laws and regulations\footnote{The entire list of statutes we selected can be found on our official GitHub.}. We then manually downloaded the most up-to-date versions of these laws from the government's official website. These laws were subsequently divided into the smallest searchable units based on articles using automated scripts. 


\subsection{Recruitment and Payment of Annotators}


For recruitment, we sourced annotators from prominent law schools\footnote{Many existing legal datasets choose law school students as annotators~\cite{li2023muser,zongyue2023leec,liu2023leveraging}.}. 
The annotation team initially consisted of 16 members. Although three members departed during the project, their positions were quickly filled to maintain the team size.
Our salary plan remunerates participants based on the number of annotations they complete, with a fixed rate of approximately 10 CNY per annotation. On average, annotators processed four queries per hour, resulting in an average hourly wage of 40 CNY. This pay rate significantly exceeds the minimum hourly wage mandated in Beijing.


\subsection{Annotation Process}

Annotators are tasked with identifying relevant articles of statutes in response to actual legal queries posed by the general public. The specifics of the annotation framework are detailed in Section ~\ref{sec:relevance}. Additionally, annotators are instructed not to use generative models, such as ChatGPT, for assistance.
The annotation process starts with the manual anonymization of each question within the STARD dataset, involving the removal of any potential identifiers associated with entities, corporations, or individuals.
Subsequently, annotators are required to locate relevant statutes for each question, following the three-step principle introduced in Section~\ref{sec:relevance}.


\subsection{Annotation Consistency}
For each question, two annotators were assigned. \textbf{The final gold standard for each question was established only when both annotators agreed on the same legal provisions\footnote{In cases where annotators had differing opinions, the question would not be included in the final dataset.}.} 

To evaluate the reliability of agreement among human annotators, we utilized Cohen's Kappa~\cite{cohen1960coefficient} $\mathcal{K}$ coefficient in a binary classification context. Each query-statute article pair corresponds to a binary classification task, where annotators judge whether the query is related or unrelated to the statute.
This analysis, conducted on a dataset comprising 1,543 annotated instances, yielded a $\mathcal{K}$ value of 0.5312. This indicates moderate agreement. 
Achieving such a $\mathcal{K}$  value is considered satisfactory for a complex task involving fifty thousand classifications with multiple possible correct labels. 

\subsection{Ethics Discussion}


We have thoroughly addressed the following ethical considerations:
\begin{itemize}[leftmargin=*]
\item \textbf{(1) Privacy and Anonymity:}
Given the sensitive nature of legal consultations, we have rigorously anonymized all queries in the STARD dataset. 
\item \textbf{(2) Transparency:}
To promote reproducibility and transparency, we have made the dataset, associated models, and the codes publicly available\footnote{https://github.com/oneal2000/STARD/tree/main}. This allows other researchers to verify, replicate, and expand upon our work, advancing the field of legal informatics.
\item \textbf{(3) Accountability:}
Recognizing the dynamic nature of legal statutes, we commit to regularly updating the STARD dataset to reflect the latest changes in law. This ensures the dataset remains accurate and reliable for ongoing research and application.
\item \textbf{(4) Accessibility:}
The STARD dataset is freely available for download from the official website under the MIT license, facilitating easy access for researchers and practitioners alike. This promotes broader usage and supports innovation across various fields.

\end{itemize}


\section{Dataset Statistics and Analysis}
The basic statistics of our proposed dataset are shown in Table~\ref{tab:basic_statistics}. 
STARD comprises a total of 1,543 queries and a large-scale corpus of 55,348 candidate statutory articles. Among these candidate statutory articles, 1,445 articles are relevant to at least one of the queries in the dataset.
The average length of a query is 27.3 words, and the average length of a statute article is nearly 120 words.

\begin{figure}[t]
\centering
    \includegraphics[width=\columnwidth]{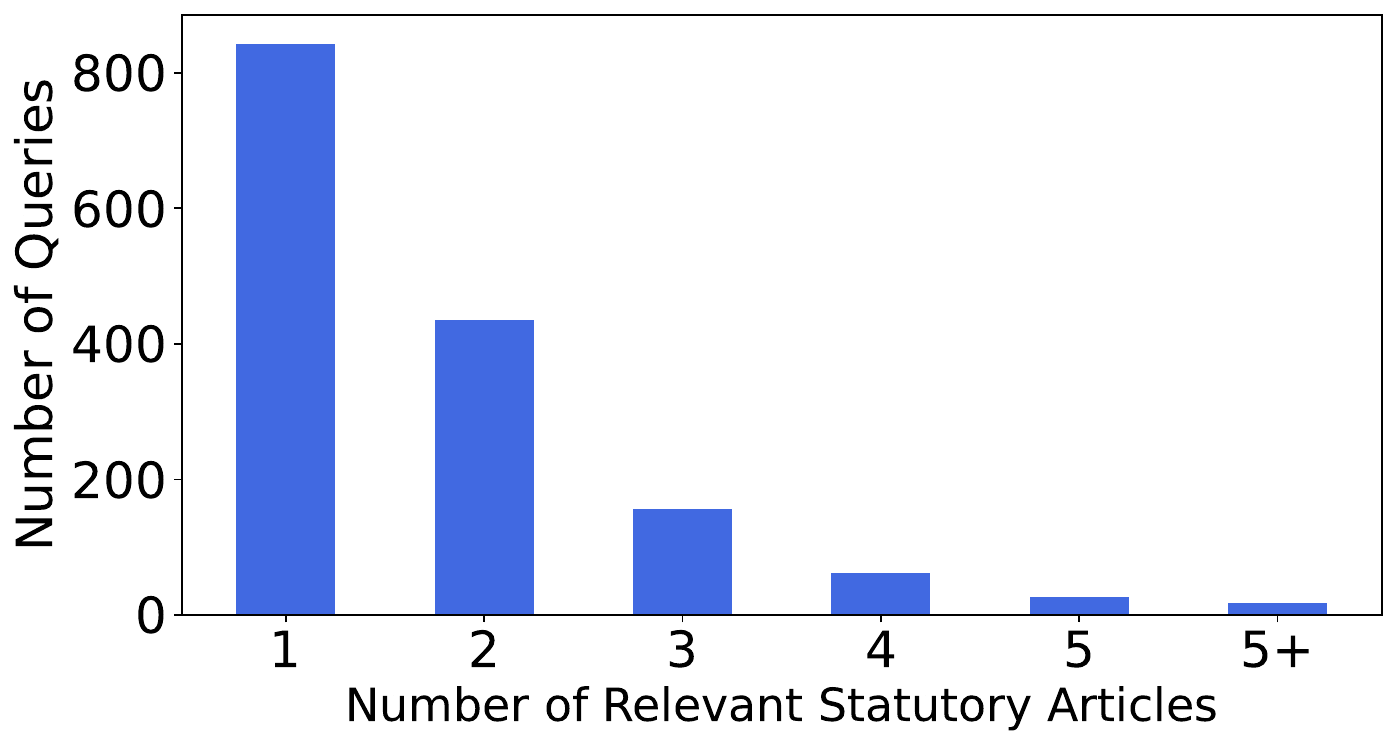}
    {\caption{Distribution of relevant statutory article numbers for each query.} \label{pic:Distribution}}
\end{figure}

Figure~\ref{pic:Distribution} presents the distribution of queries across the number of relevant statutory articles, highlighting the varied complexity within the dataset. A substantial majority of the queries, 843 out of 1,543, correspond to just one relevant statutory article, indicating a significant number of queries can be addressed with a single, specific legal reference. This could suggest that many of the non-professional queries are focused and pertain to specific legal issues that require straightforward statute retrieval.
However, 45\% of queries require multiple statutory articles which indicates some of the questions are more complex, involving multiple references of law. This diversity in query complexity demonstrates that our dataset is capable of accommodating a wide range of legal questions, from straightforward to highly intricate.

\begin{table}[t]
\centering
\caption{Basic statistics of our proposed STARD dataset.}
\label{tab:basic_statistics}
{\small{\begin{tabular}{cc}
\toprule
\textbf{Statistic}               & \textbf{\# Number} \\
\toprule
Total Queries                    & 1,543            \\
Total Candidate Statutory Articles & 55,348           \\
Total Num of Relevant Statutory Articles & 1,445           \\
Occurrences of Relevant Articles & 2,717 \\
Avg. Relevant Articles per Query & 1.76            \\
Avg. Query Length                & 27.30            \\
Avg. Article Length              & 119.93          \\
\toprule
\end{tabular}}}
\end{table}

\section{Statute Retrieval Experiment}

\begin{table*}[t]
\centering
\caption{The overall experimental results of multiple baselines on STARD. The best results are in bold, and the second-best results are underlined. ``R'' stands for Recall, and ``M'' stands for MRR. General PLM and Legal PLM are all in the zero-shot setting. Note that LSI-STARD is a classification model where each statute is treated as a unique label; we report its ranking performance based on the probability for each statute.}
\label{tab:overall}
{\small
\setlength\tabcolsep{2pt} 
\begin{tabular}{clcccccllccc}
\toprule
                                           &                        & \textbf{R@5} & \textbf{R@10} & \textbf{R@20} & \textbf{R@30} & \textbf{R@50}  & \textbf{R@100}&\textbf{R@200}&\textbf{M@3}  & \textbf{M@5}  & \textbf{M@10} \\
\toprule
\multirow{2}{*}{\textbf{Lexical Matching}}& \textbf{QL}            & 0.3363            & 0.4020              & 0.4651             & 0.4839             & 0.5537              & 0.6515&0.7224&0.3052          & 0.3167          & 0.3304          \\
                                           & \textbf{BM25}          & 0.3349            & 0.3943             & 0.4504             & 0.4773             & 0.5240               & 0.6493&0.7035&0.3176          & 0.3251          & 0.3369          \\
                                           \midrule
\multirow{3}{*}{\textbf{Open-Domain PLM}}      & \textbf{Roberta}       & 0.3216            & 0.3908             & 0.4646             & 0.5042             & 0.5715              & 0.6633&0.7351&0.2766          & 0.2905          & 0.3010           \\
           & \textbf{SEED}          & 0.2897            & 0.3555             & 0.4264             & 0.4589             & 0.4975              & 0.5626&0.6260&0.2607          & 0.2708          & 0.2816          \\
                                           & \textbf{coCondenser}   & 0.1120             & 0.1598             & 0.2223             & 0.2659             & 0.3288              & 0.4292&0.5246&0.0847          & 0.0922          & 0.1004          \\
\midrule
\multirow{2}{*}{\textbf{Legal PLM}}                                                & \textbf{SAILER}        & 0.2330             & 0.3050              & 0.3790              & 0.4286             & 0.4885              & 0.5674&0.6463&0.2006          & 0.2115          & 0.2234          \\
                                           & \textbf{Lawformer}     & 0.2411            & 0.2989             & 0.3720              & 0.4137             & 0.4733              & 0.5478&0.6309&0.2205          & 0.2313          & 0.2412          \\
\midrule

\multirow{4}{*}{\textbf{Fine-tuned PLM}}& \textbf{Dense-STARD}    & \textbf{0.5206}   & \textbf{0.6061}    & \textbf{0.7064}    & \textbf{0.7485}    & \textbf{0.8107}     & \textbf{0.9065}&\textbf{0.9531}&\textbf{0.4372} & \textbf{0.4543} & \textbf{0.4724} \\
\textbf{}  & \textbf{Dense-GPT4} & \underline{ 0.4382}      & \underline{ 0.5174}       & \underline{ 0.5961}       & \underline{ 0.6471}       & \underline{ 0.6810}         & \underline{0.7984}&\underline{0.8521}&\underline{ 0.3842}    & \underline{ 0.3948}    & \underline{ 0.4106}    \\
 & \textbf{Dense-CAIL}& 0.0887& 0.1272& 0.1832& 0.2341& 0.2712 & 0.3281&0.3819& 0.0660& 0.0719&0.0842\\
 & \textbf{LSI-STARD}& 0.1861& 0.2069& 0.2386& 0.2564& 0.3004 & 0.3410&0.3956& 0.2062& 0.2093&0.2156\\
 \toprule

\end{tabular}}
\end{table*}

\subsection{Selected Retrieval Baselines}\label{sec:baselines}

We consider four types of baselines for comparison, including traditional IR methods, pre-trained Language models on general domain data, PLMs tailored for IR, and pre-trained language models built with legal documents. \textbf{The implementation details of these baselines are provided in Appendix~\ref{appendix:imple}}

\begin{itemize}[leftmargin=*]
    \item \textbf{Traditional IR Methods}
    
    \begin{itemize}[leftmargin=*]

        \item \textbf{QL}~\cite{ponte2017language} is a language model based on Dirichlet smoothing and has good performance on retrieval tasks.
        
        \item \textbf{BM25}~\cite{robertson2009probabilistic} is an effective retrieval model based on lexical matching that achieves good performance in retrieval tasks.
        
    \end{itemize}


    \item \textbf{General Domain Pre-trained Models}
    
    \begin{itemize}[leftmargin=*]
        

        \item \textbf{Chinese-RoBERTa-WWM}~\cite{cui2021pre} is a language model pre-trained with the Whole Word Masking strategy. 

        
    
    

        
        \item \textbf{SEED}~\cite{lu2021less} is a pre-trained text encoder for dense retrieval that achieves state-of-the-art performance.

  
        \item \textbf{coCondenser}~\cite{gao2021unsupervised} is an enhanced version of Condenser\cite{gao2021condenser} that adds an unsupervised corpus-level contrastive loss to warm up the passage embedding space. 
        
    \end{itemize}

    \item \textbf{Legal Domain Pre-trained Models}
    
    \begin{itemize}[leftmargin=*]
        
        \item \textbf{Lawformer}~\cite{xiao2021lawformer} apply Longformer\cite{beltagy2020longformer} to initialize and train with the MLM task on the legal domain.

        \item \textbf{SAILER}~\cite{li2023sailer} is a structure-aware pre-trained language model for tailored legal document representation. It utilizes the logical connections between different sections within a legal document.


    \end{itemize}
    

\item\textbf{Fine-tuned Dense Retrieval Model}



    \begin{itemize}[leftmargin=*]
        
        \item \textbf{Dense-CAIL} is a dense retrieval model trained on the CAIL2018 dataset~\cite{xiao2018cail2018}. We choose this baseline to verify whether the existing dataset based on formal professional questions is sufficient for addressing statute retrieval tasks based on non-professional queries.

        \item \textbf{Dense-STARD} employs a five-fold cross-validation technique on the STARD dataset. 
        
    \end{itemize}

We initialize the above two models with Chinese-Roberta-WWM~\cite{cui2021pre}. 
For the setting of cross-validation, the dataset is randomly divided into five subsets, where one subset serves as the test set and the remaining four are used as training sets. 
The details of our fine-tuning process are introduced in Appendix~\ref{appendix:finetune}.

\item \textbf{Dense-GPT4}: We distill a dense retrieval model from GPT-4. The process involved using GPT-4 to generate legal questions based on statute articles within a given corpus. Specifically, we prompted GPT-4 to create a legal question \( q \) that is closely related to a specific statute article \( a_i^+ \), resulting in a query-statute pair \( (q, a_i^+) \). 
Then, we employ a contrastive learning approach utilizing these query-statute pairs to train the dense retriever. Details are provided in Appendix~\ref{appendix:distill}.

\item\textbf{LSI-STARD} is a Transformer based classifier fine-tuned on STARD.
In the Legal Statute Identification (LSI) field~\cite{zhong2018overview,paul2022lesicin,chalkidis2021lexglue}, the statute retrieval task is approached as a classification problem, where each statute is treated as a unique label. 
This method transforms the task into classifying legal documents or queries against a set of labels, each representing a different statute.
Following this methodology, we finetune a transformer-based classification model on the STARD dataset, employing the same five-fold cross-validation setting. We initialize the transformer-based model with Chinese-Roberta-WWM~\cite{cui2021pre} and randomly initialize the outermost MLP Layer. Details are provided in Appendix~\ref{appendix:lsi}.

\end{itemize}

\subsection{Evaluation Metrics}

We use Mean Reciprocal Rank and Recall as evaluation metrics. 
By using both MRR and Recall, we can gain insights into both the accuracy of the top-ranked results and the comprehensiveness of the relevant statutory articles retrieved by the retrieval model.
Detailed definitions of these metrics are provided in Appendix~\ref{appendix:metrics}.

\subsection{Experimental Results}

In this subsection, we provide a detailed analysis of the performance of various retrieval baselines evaluated on our proposed STARD dataset. 
We have the following insights into the effectiveness of different retrieval methods:

\textbf{(1)} Under the zero-shot setting, traditional lexical matching techniques surpass both general and legal-domain pre-trained language models (PLMs). This demonstrates that lexical matching methods are still very strong baselines in retrieval tasks. 
\textbf{(2)} Among all the methods that do not use human annotation, the performance of Dense-GPT4 stands out, exceeding that of all unsupervised methods tested. This indicates that distilling GPT4 to train task-specific models is a good choice in scenarios without human annotations. 
\textbf{(3)} Domain-specific models like SAILER are optimized for particular tasks, thus resulting in underperformance compared to general domain models. Specifically, SAILER is tailored for legal case retrieval involving long documents as queries. Consequently, it struggles with tasks that involve short queries and medium-length articles, unlike the model STARD. 
\textbf{(4)} The retrieval model fine-tuned on the CAIL2018 dataset performed sub-optimally on the STARD dataset. This suggests significant differences between the non-professional queries in STARD and the formal legal queries in existing datasets. Consequently, it underscores the unique nature of STARD, necessitating specialized models for effective statute retrieval.
\textbf{(5)} While the LSI classifier performs well in existing studies for tasks involving the classification of a few dozen statutes, it struggles with the STARD dataset, which contains over 50,000 labels, resulting in suboptimal performance. As a result, retrieval methods are more effective than the LSI approach for large-scale statute retrieval tasks.
\textbf{(6)} The performance of both the lexical matching method and the non-finetuned models is less effective than that of the Dense-STARD model. 
This arises because the former models lack the capacity to interpret life issues as legal facts, a capability that Dense-STARD has acquired through fine-tuning. It has been trained to associate the life issues presented in queries with relevant legal articles. 
However, Dense-STARD's training set is confined to just over one thousand query-article pairs. Consequently, its recall rates remain suboptimal, with Recall@100 at only 90.65\%. 
These findings underscore the necessity for further exploration in this field.


\section{Retrieval Augmented Generation Experiment}

\subsection{Selected Benchmark}
We select two datasets encompassing three tasks for our RAG experiment:

\begin{itemize}[leftmargin=*]

\item \textbf{JecQA}~\cite{zhong2020jec} is the most extensive multiple-choice dataset within the Chinese legal field. This dataset includes two distinct tasks: Knowledge-Driven Questions (KD-questions) and Case-Analysis Questions (CA-questions), encompassing a total of 26,365 questions. All the questions are multi-select, meaning that more than one option can be correct.

\item \textbf{CAIL 2018}~\cite{xiao2018cail2018} is a large-scale Chinese legal dataset designed for judgment prediction with over 2.6 million criminal cases. This dataset contains detailed annotations of judgment results, including applicable law articles, specific charges, and prescribed prison terms. We select the Charge Prediction task of CAIL 2018 and use prediction Accuracy as the evaluation metric.

\end{itemize}

\subsection{Selected LLMs and Settings}

Our selected LLMs are introduced in Appendix~\ref{appendix:LLMs}. The generation configuration is detailed in Appendix~\ref{appendix:generation}. The prompt template for LLMs is detailed in Appendix~\ref{appendix:prompt}.

\subsection{Experimental Results}

\begin{table}[t]
\caption{The overall experimental results of three LLMs on the JecQA benchmark. We report accuracy as the evaluation metric. The best results are in bold and the second best results are underlined. }
\label{tab:rag}
\centering
\setlength\tabcolsep{3pt}
{\small
\begin{tabular}{ccccc}
\toprule
                                        & \textbf{Retriever}  & \textbf{JQA-CA} & \textbf{JQA-KD}  &\textbf{CAIL}\\
                                        \toprule
\multirow{3}{*}{\textbf{Baichuan}}  & \textbf{w/o RAG}    & 0.231& 0.266&\textbf{0.850}\\
                                        & \textbf{BM25}       & \underline{0.233}& \underline{0.288}&0.766\\
                                        & \textbf{Dense-STARD} & \textbf{0.238}& \textbf{0.291}&\underline{0.816}\\
                                        \midrule
\multirow{3}{*}{\textbf{chatGLM}}   & \textbf{w/o RAG}    & 0.185& 0.194&0.636\\
                                        & \textbf{BM25}       & \underline{0.189}& \underline{0.224}&\underline{0.646}\\
                                        & \textbf{Dense-STARD} & \textbf{0.200}& \textbf{0.237}&\textbf{0.684}\\
                                        \midrule
\multirow{3}{*}{\textbf{chatGPT}} & \textbf{w/o RAG}    & 0.187& 0.206&0.496\\
                                        & \textbf{BM25}       & \textbf{0.233}& \textbf{0.293}&\textbf{0.528}\\
                                        & \textbf{Dense-STARD} & \underline{0.193}& \underline{0.252}&\underline{0.503}\\
                                        \toprule
\end{tabular}}
\end{table}

Table~\ref{tab:rag} presents the results of the LLM's performance with and without the use of Retrieval-Augmented Generation (RAG). In the scenario without RAG, the LLM directly outputs the correct options based on the question. In the RAG scenario, the retrieval model (BM25 or Dense-STARD) recalls the top 10 relevant statutory articles from the corpus based on the question. The retrieved statutory articles are then integrated into a meticulously designed prompt template (detailed in Appendix~\ref{appendix:prompt}).

The experimental results reveal that using the STARD corpus as the external knowledge base for the RAG significantly enhances the performance of large language models (LLMs) and underscores the value of our proposed dataset in improving the effectiveness of LLMs on legal tasks. 
The results also reveal that different LLMs have unique preferences for retrievers. 
For the Baichuan and ChatGLM models, a fine-tuned dense retriever surpasses BM25, indicating that these models benefit from dense retrievers' high recall rates. 
However, this advantage is not observed with the ChatGPT model, where BM25 outperforms the fine-tuned dense retriever. 
This suggests that the performance of RAG is highly dependent on the preferences of the LLM regarding the retriever.
The experimental results on the CAIL 2018 dataset align with those observed for JecQA, with one notable exception: the performance of the Baichuan model without RAG. In this setting, Baichuan's performance is markedly superior to that of chatGLM, chatGPT, and Baichuan with RAG. We hypothesize that this exception arises from the Baichuan model's utilization of the CAIL 2018 dataset during its pre-training phase, leading to a direct answer accuracy rate that is even 81\% higher than that of chatGPT.

\section{Related Work}
CAIL 2018~\cite{xiao2018cail2018, zhong2018overview} competitions conduct law statute retrieval work using formal legal judgment documents. The queries in the dataset originate from the ``Court's Findings'' part of the judgments, and the candidates are statute articles of Chinese Criminal Law. 
The annual COLIEE competitions introduce a series of statute retrieval datasets using the questions extracted from the Japanese legal bar exams~\cite{goebel2023summary,kim2022coliee,rabelo2022overview,li2023thuir,li2023thuir2}. 
These tasks aim to retrieve relevant statute law from the Japanese Civil Code Article according to questions from bar exams.
AILA~\cite{bhattacharya2019fire} competitions also introduce a series of statute retrieval datasets. 
The queries from AILA are legal judgment documents from the Supreme Court of India. 
The candidate statutes are part of the set of statute articles from Indian law.
BSARD~\cite{louis2021statutory} is a statutory article retrieval dataset in French with candidate articles from a 22,600+ Belgian law articles corpus. 

In the studies of the Legal Statute Identification (LSI)~\cite{zhong2018overview,paul2022lesicin,chalkidis2021lexglue}, finding the relevant statute is approached as a classification problem, where each statute is treated as a unique label. 
LADAN~\cite{xu2020distinguish} is an LSI method that uses a graph neural network and attention mechanism to distinguish confusing law articles.
LeSICiN~\cite{paul2022lesicin} utilizes both textual content and legal citation networks to identify relevant legal statutes.

Legal QA tasks also aim to fulfill the public’s demand for legal information~\cite{do2017legal}.
LLeQA~\cite{louis2024interpretable} is a French long-form legal QA dataset comprising 1,868 expert-annotated legal questions.
GerLayQA~\cite{buttner2024answering} is a question-answering dataset comprising 21k laymen’s legal questions paired with answers from lawyers and grounded in concrete law book paragraphs.





\section{Conclusion}
We present STARD, a new benchmark consisting of 1,543 questions from the general public. To the best of our knowledge, STARD is the first Chinese statutes retrieval dataset tailored for the general public. 
Moreover, we propose an annotation framework to improve the accuracy and relevance of statute retrieval annotation, which offers valuable guidelines for future legal annotations. 
Our experiments across various retrieval models highlighted the complexities of non-professional statute retrieval, indicating the necessity for further exploration. Additionally, we demonstrated that integrating the STARD dataset significantly boosts the performance of LLMs in legal tasks, showcasing its potential to enhance legal AI applications.

\section{Limitations}
We acknowledge the limitations of this paper. One of the primary limitations is that our dataset is specifically designed around the Chinese legal system, inherently limiting its direct applicability to legal systems outside of this context. Despite our discussions on potential methodologies for adapting STARD to other civil law systems, such an expansion necessitates creating and annotating new datasets tailored to those systems' distinct legal frameworks and statutes. Thus, our future work will be dedicated to developing additional datasets that encompass a broader range of civil law systems. This endeavor aims to extend the utility of our work and foster further research and development in the domain of legal statute retrieval, ensuring broader applicability and relevance across different legal landscapes.

\section{Ethics Statement}
In the framework of this research, ethical considerations have been paramount from the initial stages, underscoring our commitment to the responsible advancement and application of artificial intelligence technologies. Our adherence to the principles of open research and the critical importance of reproducibility have compelled us to make all associated models, datasets, and codebases publicly available on GitHub. 

Moreover, in the development of our dataset, we have paid scrupulous attention to privacy and respect for individuals' rights. Given the inherently sensitive nature of legal consultations, we have diligently anonymized every query within the STARD dataset. This process involved the removal of any potential identifiers related to entities, corporations, or individuals, thereby safeguarding privacy and preempting the possibility of data misuse. 

\bibliography{custom}
\appendix

\section{License and Permissions}
\label{appendix:license}
STARD is available under the MIT License. This permissive license was chosen to encourage the widespread use and adaptation of our resources, allowing for both academic and commercial applications without significant restrictions. For detailed terms and conditions, including how the dataset, code, and models can be used, modified, and shared, please refer to the documentation provided in our GitHub repository\footnote{https://github.com/oneal2000/STARD/tree/main}.

\section{Implementation Details of Retrieval Baselines}
\label{appendix:imple}
\begin{itemize}[leftmargin=*]

\item For the implementation of traditional IR methods QL and BM25, we use the Pyserini toolkit: {https://github.com/castorini/pyserini}.

\item For the implementation of Chinese-RoBERTa-WWM, we directly use their models released on Huggingface\footnote{https://huggingface.co/hfl/chinese-roberta-wwm-ext}.
As SEED and Condenser have no available Chinese versions, we reproduce their work on the Chinese Wikipedia based on their open-source training code and follow all settings provided in their paper~\cite{lu2021less,gao2021condenser}.

\item For the implementation of {Lawformer}~\cite{xiao2021lawformer} and {SAILER}~\cite{li2023sailer}, we directly use the checkpoints released on the official GitHub\footnote{https://github.com/CSHaitao/SAILER/, https://github.com/thunlp/LegalPLMs}. 

\end{itemize}

\section{Evaluation Metrics}
\label{appendix:metrics}
Following the setting of previous works~\cite{li2023towards,chen2023thuir,su2023thuir2,ma2023caseencoder,chen2022web,ye2024relevance}, we use Mean Reciprocal Rank and Recall as evaluation metrics.

The Mean Reciprocal Rank is a statistical measure used to evaluate the performance of a query-based system, where the primary goal is to retrieve the highest-ranked item. MRR calculates the average of the reciprocal ranks of results for a sample of queries. The reciprocal rank of a query response is the multiplicative inverse of the rank of the first correct answer:


{
\begin{equation}
\text{MRR} = \frac{1}{Q} \sum_{i=1}^Q \frac{1}{\text{rank}_i}
\end{equation}
}

\noindent where \(Q\) is the number of queries, and \(\text{rank}_i\) is the rank position of the first relevant document for the \(i\)-th query.

Recall measures the ability of a model to retrieve all relevant instances in a dataset. It is defined as the ratio of the number of relevant items correctly retrieved to the total number of relevant items in the database, which is critical in scenarios where missing any relevant item could be costly:

\begin{equation}
\text{Recall} = \frac{\text{Number of relevant items retrieved}}{\text{Total number of relevant items}}
\end{equation}

\section{Selected LLMs}
\label{appendix:LLMs}

Our selected LLMs are listed as follows:
\begin{itemize}[leftmargin=*]

\item \textbf{Baichuan}~\cite{yang2023baichuan} is a series of large-scale multilingual language models, trained from scratch on 2.6 trillion tokens. We choose the \textbf{Baichuan-2-Base-13B} model which is widely used in bilingual Chinese-English scenarios.

\item \textbf{ChatGLM}~\cite{du2022glm} is a series of generative language models optimized for Chinese question answering and dialogue. We choose \textbf{ChatGLM3-6B} with 6.2 billion parameters.

\item \textbf{ChatGPT} ~\cite{brown2020language} is a series of large language models developed by OpenAI, including several versions. Among these, we choose \textbf{GPT-3.5-turbo}, which is identified as the most advanced GPT-3.5 model.

\end{itemize}

\section{Generation Configuration}
\label{appendix:generation}
We obtain responses from chatGPT by accessing its official API~\footnote{https://platform.openai.com/docs/guides/text-generation/chat-completions-api}. For Baichuan and chatGLM, we directly download model parameters from each model's official Hugging Face repositories and use the official Python code provided by Hugging Face to obtain the response. We use the official default configurations provided by each model for the generation configuration.

\section{Prompt Template for RAG}
\label{appendix:prompt}
Previous studies have shown that RAG can significantly improve the performance of LLMs~\cite{lewis2020retrieval,su2024dragin} and mitigate the hallucination phenomenon in LLMs~\cite{zhang2023siren,su2024unsupervised}.
In our RAG experiments, we employed the following prompt template for the LLM: 

\begin{tcolorbox}[colback=lightgray!20,colframe=darkgray!80,title=Prompt 1]

Please answer the question based on the following statute articles:

\textbf{Article 1}:  [Content]
   
   ......
   
\textbf{Article 10}: [Content]
   
Please answer the following question based on the provided articles and your knowledge, prioritizing the provided knowledge. Note that the provided articles might not include those relevant to the question. 

\textbf{Question:} xxx

\end{tcolorbox}

\section{Fine-tuning Process}
\label{appendix:finetune}

We initialize the model with Chinese-Roberta-WWM~\cite{cui2021pre}.
We use the dual-encoder architecture~\cite{karpukhin2020dense,fang2024scaling,su2023caseformer} to compute the dot product between two embedding vectors as the relevance score:

{\small
\begin{equation}
\label{cross_input_dual}
X(c) = [CLS] q [SEP] ,
\end{equation}
}

{\small{\begin{equation}
\label{cross_input_dual}
X(s) = [CLS] s [SEP] ,
\end{equation}}
}

{\small
\begin{equation}
\label{embedding}
Emb(X) =  transformer_{[CLS]}(X) ,
\end{equation}
}

{\small
\begin{center}
\begin{equation}
\label{score_dense}
\begin{split}
S(q, s) = Emb(X(q))^\top \cdot Emb(X(s)) ,
\end{split}
\end{equation}
\end{center}
}

\noindent where $q$ is the query, $s$ is the statute, $transformer_{[CLS]}(\cdot)$ outputs a contextualized vector for each token and we select the "[CLS]" vector as the embedding vector of the input. In Equation ~\ref{score_dense}, we regard the inner products of embeddings as the relevance score $S$.

For the loss function, we use the Softmax Cross Entropy Loss~\cite{cao2007learning,ai2018learning,gao2021rethink,su2024wikiformer} to optimize the retrieval model, which is defined as:

\begin{equation}
    \label{eq:LRP}
\small{\begin{aligned}  
   & \mathcal{L}(Q,s^+,N) \\& = -\log_{}{    \frac{exp(S(Q,s^+))}{exp(S(Q,s^+) + \sum_{s^-\in N} exp(S(Q,s^-))}} ,
\end{aligned}}
\end{equation}

\noindent where $S$ is the relevance score function which is defined in Equation~\ref{score_dense}. $Q$ is the query, $s^+$ is the relevant statute and $N$ is the set of irrelevant statutes randomly sampled from the corpus.

\section{Training Process of the Dense Retrieval Model Distilled from GPT-4}
\label{appendix:distill}
We introduce an approach utilizing GPT-4 to generate labels for question-article pairs. Our methodology leverages GPT-4's capabilities to autonomously generate non-professional legal questions from statutory articles, thus enabling the pairing of these questions with their corresponding articles without the need for human supervision.

The process begins by selecting statutory articles from the corpus of STARD. GPT-4 is then tasked with generating a legal question based on the content of each article. This is achieved by providing GPT-4 with a specific prompt designed to simulate a scenario in which an individual without prior legal knowledge seeks advice. The prompt instructs GPT-4 to formulate a question that such an individual might ask, ensuring that the question is directly related to and explainable by the content of the statutory article provided. The prompt used in this study is structured as follows:

\begin{tcolorbox}[colback=lightgray!20,colframe=darkgray!80,title=Prompt 1]
Given the following known statutory article:\\

\textbf{[Content of the statutory article]}\\

Imagine a scenario in which a person without legal knowledge is seeking legal advice. Please generate a question that this party might ask.\\

Note: The question must be fully explainable using the statutory article mentioned above, and remember that the person who proposes this question has never read the legal articles mentioned before.
\end{tcolorbox}

\noindent Each interaction with GPT-4 results in the creation of a query-statute pair \((q, a_i^+)\), where \(q\) is the generated question and \(a_i^+\) is the positive statute article to which the question is relevant.

Following the generation of query-statute pairs, we employ a contrastive learning framework to train a dense retriever model. We use the same relevance scoring function \(S\), as detailed in Equation~\ref{score_dense}, which assesses the relevance of articles to the questions.

In the training phase, for each query \(Q\) paired with a positive article \(a_i^+\), we also sample \(8\) negative articles from the corpus. These negative samples are not relevant to the query and serve as the negative set. The loss function employed, represented by Equation~\ref{eq:LRP}, is designed to maximize the score of the positive article relative to the scores of the negative samples, effectively training the model to distinguish between relevant and non-relevant articles accurately.

\begin{equation}
    \label{eq:LRP}
\small{\begin{aligned}  
   & \mathcal{L}(Q,a_i^+,N) \\& = -\log_{}{    \frac{exp(S(Q,a_i^+))}{exp(S(Q,a_i^+) + \sum_{a^-\in N} exp(S(Q,a^-))}} ,
\end{aligned}}
\end{equation}

\noindent where $S$ is the relevance score function, which is defined in Equation~\ref{score_dense}, and $N$ is the set of irrelevant statutes randomly sampled from the corpus.

\section{Training Process of the LSI Classifier}
\label{appendix:lsi}

We apply a fine-tuned classifier approach to evaluate the performance of Legal Statute Identification (LSI) methods, as defined in previous works~\cite{zhong2018overview,paul2022lesicin}. LSI is framed as a classification task where each legal statute is treated as a distinct label. This transformation allows for the classification of legal documents or queries by associating them with the relevant statutory labels.

Our methodology utilizes a transformer-based classification model, specifically fine-tuned on the STARD dataset within a five-fold cross-validation framework. We initiate our model using the Chinese-Roberta-WWM~\cite{cui2021pre} for the transformer's parameters, while the parameters for the outermost Multi-Layer Perceptron (MLP) layer are initialized randomly. The process of input transformation and subsequent classification is defined by the following equations:

{\small{\begin{equation}
X(q) = [CLS] q [SEP],
\end{equation}

\begin{equation}
L(q) = MLP(transformer_{[CLS]}(X(q))),
\end{equation}}}

\noindent where \( q \) represents a query from the STARD dataset. The function \( transformer_{[CLS]}(\cdot) \) first encodes the input using the transformer architecture, focusing on the output of the [CLS] token's embedding vector. The \( MLP(\cdot) \) then maps this embedding onto the space of statutory labels \( L \).

\end{document}